\documentclass[preprint, 3p, sort&compress]{elsarticle}
\usepackage{dcolumn}
\usepackage{booktabs,tabularx} 
\usepackage{booktabs}
\usepackage{textcomp}
\usepackage{amssymb}
\usepackage{amsthm}
\usepackage[mathlines]{lineno}
\usepackage{graphicx}
\usepackage{units}
\usepackage{url}
\usepackage{amsmath}
\usepackage{amsfonts}
\usepackage{bm}
\usepackage{textcomp}
\usepackage{subfigure}
\usepackage{multicol}
\usepackage{multirow}
\usepackage{verbatim}
\usepackage{rotating}
\usepackage[colorlinks,linkcolor=black,citecolor=red]{hyperref}
\usepackage{float}
\usepackage{epsfig}
\usepackage{dcolumn}
\usepackage{bm}
\usepackage{color}
\usepackage{pstricks}
\usepackage{pst-node}
\usepackage{times}
\usepackage{indentfirst}
\usepackage[english]{babel}
\usepackage{epstopdf}
\addto{\captionsenglish}{%

}
\journal{Physics Letters B}


\usepackage[font=footnotesize]{caption}
\usepackage{overpic}

\newcommand{\mevcc}{\,\mbox{MeV}/c^2}
\newcommand{\gev}{\,\mbox{GeV}}
\newcommand{\gevc}{\,\mbox{GeV}/c}
\newcommand{\gevcc}{\,\mbox{GeV}/c^2}
\newcommand{\mgg}{M_{\gamma\gamma}}

\biboptions{numbers,sort&compress}
\begin{document}
\begin{frontmatter}
\title{\boldmath \bf Search for an axion-like particle in radiative $J/\psi$ decays}
\author{
\small
M.~Ablikim$^{1}$, M.~N.~Achasov$^{13,b}$, P.~Adlarson$^{73}$, R.~Aliberti$^{34}$, A.~Amoroso$^{72A,72C}$, M.~R.~An$^{38}$, Q.~An$^{69,56}$, Y.~Bai$^{55}$, O.~Bakina$^{35}$, R.~Baldini Ferroli$^{28A}$
% ...
\\
\vspace{0.2cm}
(BESIII Collaboration)\\
\vspace{0.2cm}  
$^{1}$ Institute of High Energy Physics, Beijing 100049, People's Republic of China\\
% ...
$^{13}$ G.I. Budker Institute of Nuclear Physics SB RAS (BINP), Novosibirsk 630090, Russia\\
% ...
$^{28}$ INFN Laboratori Nazionali di Frascati , (A)INFN Laboratori Nazionali di Frascati, I-00044, Frascati, Italy; (B)INFN Sezione di  Perugia, I-06100, Perugia, Italy; (C)University of Perugia, I-06100, Perugia, Italy\\
% ...
$^{34}$ Johannes Gutenberg University of Mainz, Johann-Joachim-Becher-Weg 45, D-55099 Mainz, Germany\\
$^{35}$ Joint Institute for Nuclear Research, 141980 Dubna, Moscow region, Russia\\
% ...
$^{38}$ Liaoning Normal University, Dalian 116029, People's Republic of China\\
% ...
$^{55}$ Southeast University, Nanjing 211100, People's Republic of China\\
$^{56}$ State Key Laboratory of Particle Detection and Electronics, Beijing 100049, Hefei 230026, People's Republic of China\\
% ...
$^{69}$ University of Science and Technology of China, Hefei 230026, People's Republic of China\\
% ...
$^{72}$ University of Turin and INFN, (A)University of Turin, I-10125, Turin, Italy; (B)University of Eastern Piedmont, I-15121, Alessandria, Italy; (C)INFN, I-10125, Turin, Italy\\
$^{73}$ Uppsala University, Box 516, SE-75120 Uppsala, Sweden\\
% ...
$^{b}$ Also at the Novosibirsk State University, Novosibirsk, 630090, Russia\\
% ...
\vspace{-1.5cm}
}
\date{\today}
\begin{abstract}
We search for an axion-like particle (ALP) $a$ through the process $\psi(3686)\rightarrow\pi^+\pi^-J/\psi$, $J/\psi\rightarrow\gamma a$, $a\rightarrow\gamma\gamma$ in a data sample of $(2.71\pm0.01)\times10^9$ $\psi(3686)$ events collected by the BESIII detector. No significant ALP signal is observed over the expected background, and the upper limits on the branching fraction of the decay $J/\psi\rightarrow\gamma a$ and the ALP-photon coupling constant $g_{a\gamma\gamma}$ are set at 95\% confidence level in the mass range of $0.165\leq m_a\leq2.84\gevcc$. The limits on $\mathcal{B}(J/\psi\rightarrow\gamma a)$ range from $8.3\times10^{-8}$ to $1.8\times10^{-6}$ over the search region, and the constraints on the ALP-photon coupling are the most stringent to date for $0.165\leq m_a\leq1.468\gevcc$.

\end{abstract}
\end{frontmatter}

\begin{multicols}{2}

\section{Introduction}

Axion-like particles (ALPs) are pseudo-Goldstone bosons arising from some spontaneously broken global symmetry, addressing the strong \emph{CP}~\cite{axion_1,axion_2,axion_3,axion_4} or hierarchy problems~\cite{hierarchy}. ALPs could appear in theories beyond the Standard Model (SM), such as string theory~\cite{string} and extended Higgs models~\cite{Higgs}. ALPs could also provide a portal connecting SM particles to the dark sectors~\cite{dark_1}, and in certain situations, they are proposed as cold dark matter candidates~\cite{dark_2,dark_3,dark_4}.
In the most common scenarios, the ALP $a$ predominantly couples to photons with a photon coupling constant $g_{a\gamma\gamma}$. 
As a generalization of QCD axions, ALPs have arbitrary masses and couplings which are bounded by experiments. In the sub-$\mbox{MeV}/c^2$ mass range, the bounds on $g_{a\gamma\gamma}$ are provided by laser experiments, solar photon instruments, as well as cosmological and astrophysical observations~\cite{cosmo_1,cosmo_2}. 
In the $\mbox{MeV}/c^2$ to $\mbox{GeV}/c^2$ mass region, constraints on the photon coupling constant mainly come from beam-dump experiments~\cite{PrimEx,Beam_dump,NA64} and high-energy collider experiments. Long-lived ALPs have been searched for in $J/\psi$ and $\Upsilon$ radiative decays~\cite{meson1,meson2} and through $e^+e^-\rightarrow\gamma+\text{invisible}$ processes~\cite{BaBar}. The limits for short-lived ALPs decaying to two photons are obtained from $e^+e^-\rightarrow\gamma\gamma$~\cite{OPAL}, $e^+e^-\rightarrow\gamma\gamma\gamma$~\cite{L3,Belle} via ALP-strahlung production, and the light-by-light scattering process $\gamma\gamma\rightarrow\gamma\gamma$~\cite{CMS,ATLAS}, as well as $pp$ collisions~\cite{collider,pp1,pp2}. There is a striking gap in the limits on the ALP-photon coupling in the ALP mass range from $100\mevcc$ to roughly $10\gevcc$, leaving the large area $10^{-3}\lesssim g_{a\gamma\gamma}\lesssim10^{-5}\,\mbox{GeV}^{-1}$ still uncovered. Electron-positron colliders can play a unique role  in further exploring this region~\cite{LHC}.

We search for ALPs decaying into two photons at BESIII in $J/\psi$ radiative decays via $J/\psi\rightarrow\gamma a$, $a\rightarrow\gamma\gamma$ in the mass range of $0.165\leq m_a\leq2.84\gevcc$. In this Letter we ignore the ALP coupling to c-quarks~\cite{Merlo} and assume the branching fraction of $a$ decaying to photons is 100\%, with a decay width $\Gamma_a=g_{a\gamma\gamma}^2m_a^3/64\pi$. The experimental signature of the ALP decaying into two photons depends on the correlation between its mass and coupling. Except for the mass region below $0.1\gevcc$, the ALP has a negligible lifetime and decay width, and the opening angle between the decay photons is large enough to be resolved in the detector. The mass intervals of $0.10<m_a<0.165\gevcc$, $0.46<m_a<0.60\gevcc$ and $0.90<m_a<1.01\gevcc$ are excluded from the search due to the peaking backgrounds from the decays of $\pi^0$, $\eta$ and $\eta'$ mesons, respectively.
In order to avoid pollution from non-resonant ALP production $e^+e^-\rightarrow\gamma a$ and to probe resonant ALP production $J/\psi\rightarrow\gamma a$ in a model-independent way~\cite{Merlo}, the decay $\psi(3686)\rightarrow\pi^+\pi^-J/\psi$ is exploited to isolate the $J/\psi$ sample.
\vspace{-0.2cm}

\section{Detector, data sets and Monte Carlo simulation}

The BESIII detector~\cite{Ablikim:2009aa} records symmetric $e^+e^-$ collisions provided by the BEPCII storage ring~\cite{Yu:IPAC2016-TUYA01}, which operates in the center-of-mass energy range from 2.00 to 4.95~GeV, with a peak luminosity of $1\times10^{33}$~cm$^{-2}$s$^{-1}$ achieved at $\sqrt{s}$ = 3.77~GeV. BESIII has collected large data samples in this energy region~\cite{Ablikim:2019hff}. The cylindrical core of the BESIII detector covers 93\% of the full solid angle and consists of a helium-based multilayer drift chamber~(MDC), a plastic scintillator time-of-flight system~(TOF), and a CsI(Tl) electromagnetic calorimeter~(EMC), which are all enclosed in a superconducting solenoidal magnet providing a 1.0~T magnetic field. The solenoid is supported by an octagonal flux-return yoke with resistive plate counter muon identification modules interleaved with steel. 
The charged-particle momentum resolution at $1~{\rm GeV}/c$ is $0.5\%$, and the ${\rm d}E/{\rm d}x$ resolution is $6\%$ for electrons from Bhabha scattering. The EMC measures photon energies with a resolution of $2.5\%$ ($5\%$) at $1$~GeV in the barrel (end-cap) region. The time resolution in the TOF barrel region is 68~ps, while that in the end-cap region is 110~ps. The end-cap TOF system was upgraded in 2015 using multigap resistive plate chamber technology, providing a time resolution of 60~ps~\cite{etof}.

We perform the search using the $\psi(3686)$ data samples collected by the BESIII detector during three data-taking periods: 2009, 2012, and 2021. The numbers of $\psi(3686)$ events are determined by counting inclusive hadronic events whose branching fraction is known rather precisely~\cite{number,number2}. There are $(107.0\pm0.8)\times10^{6}$ $\psi(3686)$ events collected in 2009, $(341.1\pm2.9)\times10^6$ events collected in 2012 and $(2.26\pm0.01)\times10^9$ events collected in 2021, for a total of $(2.71\pm0.01)\times10^9$ $\psi(3686)$ events. 
In this analysis, the $J/\psi$ sample originates from the decay $\psi(3686)\rightarrow\pi^+\pi^-J/\psi$. The analysis strategy is to first tag $J/\psi$ events by selecting two oppositely charged pions, and then to search for the signal candidate events with three additional photons originating from the tagged $J/\psi$ sample. A semi-blind procedure is performed to avoid possible bias, where approximately 10\% of the full data set is used to optimize the event selection and validate the fit approach. The final result is then obtained with the full data set only after the analysis strategy is fixed.

Simulated data samples produced with a {\sc geant4}-based~\cite{geant4} Monte Carlo (MC) package, which includes the geometric description of the BESIII detector and the detector response, are used to determine detection efficiencies and to estimate backgrounds. The simulation models the beam-energy spread and initial-state radiation in $e^+e^-$ annihilations with the generator {\sc kkmc}~\cite{ref:kkmc}. 
An inclusive MC sample with approximately the same number of $\psi(3686)$ decays as in data is used to check for potential backgrounds. The inclusive MC sample includes the production of the $\psi(3686)$ resonance, the initial-state radiation production of the $J/\psi$, and the continuum processes incorporated in {\sc kkmc}~\cite{ref:kkmc}. All particle decays are modeled with {\sc evtgen}~\cite{ref:evtgen} using branching fractions either taken from the Particle Data Group~(PDG)~\cite{pdg}, when available, or otherwise estimated with {\sc lundcharm}~\cite{ref:lundcharm}. Final-state radiation from charged final-state particles is incorporated using the {\sc photos} package~\cite{photos}. 
To simulate the signal, the decay $\psi(3686)\rightarrow\pi^+\pi^-J/\psi$ is modeled according to the partial wave analysis results of $\psi(3686)\rightarrow\pi^+\pi^-J/\psi$~\cite{jpipi}; a $P$-wave model~\cite{ref:evtgen} is used for the subsequent decay $J/\psi\rightarrow\gamma a$ and a phase-space model is used for the $a\rightarrow\gamma\gamma$ decay. The signal events are simulated individually for different mass hypotheses of ALP $m_a$ in the range of $0.165\leq m_a\leq2.84\gevcc$ with a step size of $5\mevcc$, which is smaller than the signal resolution. 
Study of the inclusive MC sample with a generic event type analysis tool, TopoAna~\cite{topoana}, indicates that the dominant backgrounds are from $\psi(3686)\rightarrow\pi^+\pi^-J/\psi$ with subsequent decays of $J/\psi\rightarrow\gamma\pi^0$, $\gamma\eta$ and $\gamma\eta'$. These backgrounds are each generated exclusively with the angular distribution of $1+\cos^2\theta_\gamma$~\cite{ref:evtgen}, where $\theta_\gamma$ is the angle of the radiative photon relative to the positron beam direction in the $J/\psi$ rest frame. Some potential backgrounds of the form $\psi(3686)\rightarrow\pi^+\pi^-J/\psi$ with $J/\psi$ decaying into purely neutral particles in the final states are generated exclusively with different generators: $J/\psi\rightarrow\gamma\eta_c$ with the angular distribution of $1+\cos^2\theta_\gamma$~\cite{ref:evtgen}, $J/\psi\rightarrow\gamma\pi^0\pi^0$ according to the partial wave analysis results of $J/\psi\rightarrow\gamma\pi^0\pi^0$~\cite{pwa_gpipi}, as well as $J/\psi\rightarrow\gamma\gamma\gamma$ with a phase-space distribution. Each background MC sample is weighted to match the integrated luminosity of the data set.
\vspace{-0.2cm}

\section{Event selection}

Signal candidates are required to have two oppositely charged tracks and at least three photon candidates. Charged tracks detected in the MDC are required to be within a polar angle ($\theta$) range of $|\rm{cos\theta}|<0.93$, where $\theta$ is defined with respect to the $z$-axis, which is the symmetry axis of the MDC. For each charged track, the momentum must be less than $0.45\gevc$ and the distance of closest approach to the interaction point (IP) along the $z$-axis $|V_{z}|$ must be less than 10\,cm, and in the transverse plane $|V_{xy}|$ less than 1\,cm. Both charged tracks are assumed to be pion candidates, and the recoil mass in the center-of-mass system $M(\pi^+\pi^-)_{\rm recoil}$ must be in the range of [3.080, 3.114]~$\gevcc$.

Photon candidates are identified using showers in the EMC. The deposited energy of each shower must be more than 25~MeV in the barrel region ($|\!\cos\theta|<0.80$) and more than 50~MeV in the end-cap region ($0.86<|\!\cos\theta|<0.92$).  To exclude showers that originate from charged tracks, the angle subtended by the EMC shower and the position of the closest charged track at the EMC must be greater than 10 degrees as measured from the IP. To suppress electronic noise and showers unrelated to the event, the difference between the EMC time and the event start time is required to be within [0, 700]~ns. Events satisfying the above requirements are retained for further analysis.

\begin{figure}[H]
\vspace{-0.4cm}
\begin{center}
    \begin{overpic}[width=8cm]{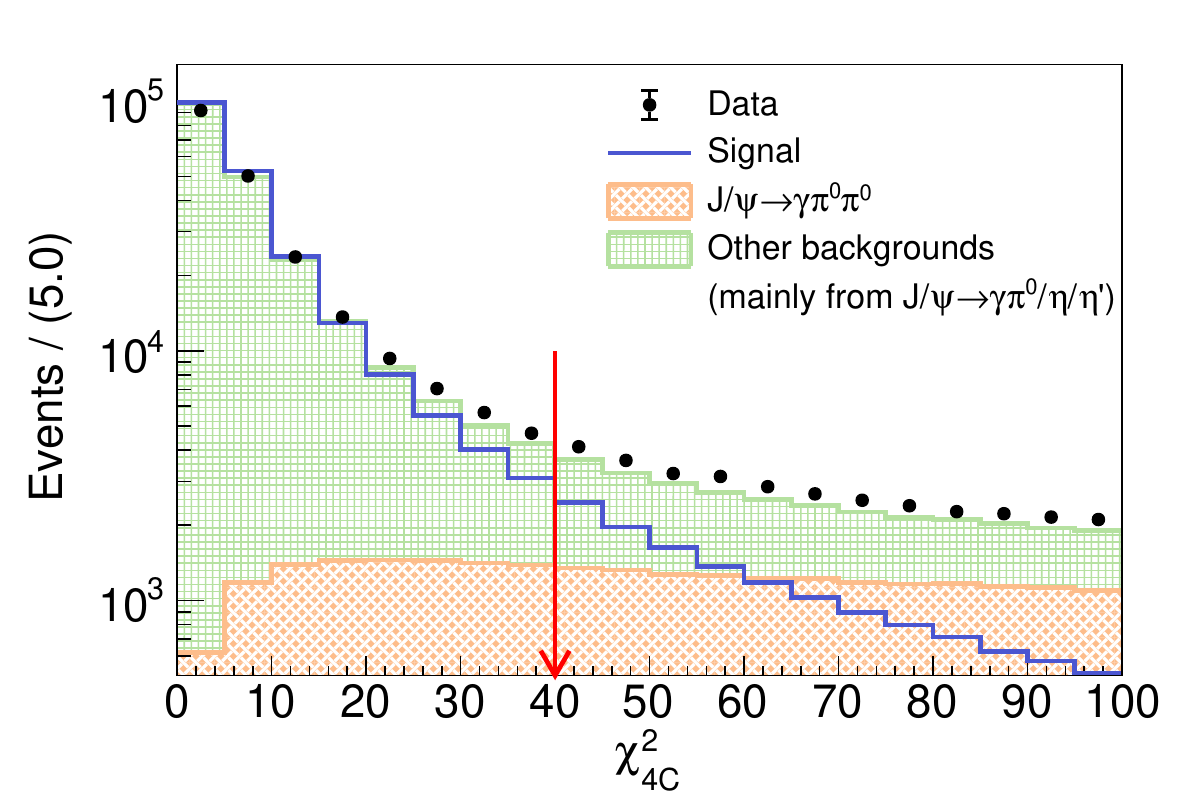}
    \end{overpic}
\end{center}
\vspace{-0.6cm}
\caption{The distribution of $\chi_{\rm 4C}^2$ for data, signal and MC-simulated backgrounds. The requirement of $\chi_{\rm 4C}^2<40$ is applied to suppress the background from $J/\psi\rightarrow\gamma\pi^0\pi^0$ decays.}
\vspace{-0.2cm}
\label{fig:chi2}
\end{figure}

The $\pi^+$ and $\pi^-$ tracks are constrained to a common vertex to determine the event interaction point, and a four-constraint~(4C) kinematic fit to the initial four-momentum of the $\psi(3686)$ is applied for all possible $\pi^+\pi^-\gamma\gamma\gamma$ combinations. The combination with the smallest fit $\chi_{\rm 4C}^2$ is retained, with the requirement of $\chi_{\rm 4C}^2<40$. The distribution of $\chi_{\rm 4C}^2$ is shown in Fig.~\ref{fig:chi2}. To further reject background from $J/\psi\rightarrow\gamma\pi^0\pi^0$ decays, we form all $\pi^+\pi^-5\gamma$ combinations if 5 or more photons are present, and we require the minimum $\chi^2_{\rm 4C}(5\gamma)$ of the corresponding kinematic fits to satisfy $\chi^2_{\rm 4C}<\chi^2_{\rm 4C}(5\gamma)$. For the remaining photons, the sum of their energy deposition in the EMC $E_{\rm other}$ is required to be less than $0.1\gev$. The requirements of $\chi_{\rm 4C}^2$ and $E_{\rm other}$ are optimized according to the Punzi significance~\cite{punzi} defined as $\epsilon/(1.5+\sqrt{B})$, where $\epsilon$ denotes the signal efficiency obtained from signal MC samples and $B$ is the number of background events obtained from background MC samples. 

\vspace{-0.2cm}

\section{ALP signal search}

\begin{figure}[H]
\vspace{-0.4cm}
\begin{center}
    \begin{overpic}[width=8cm]{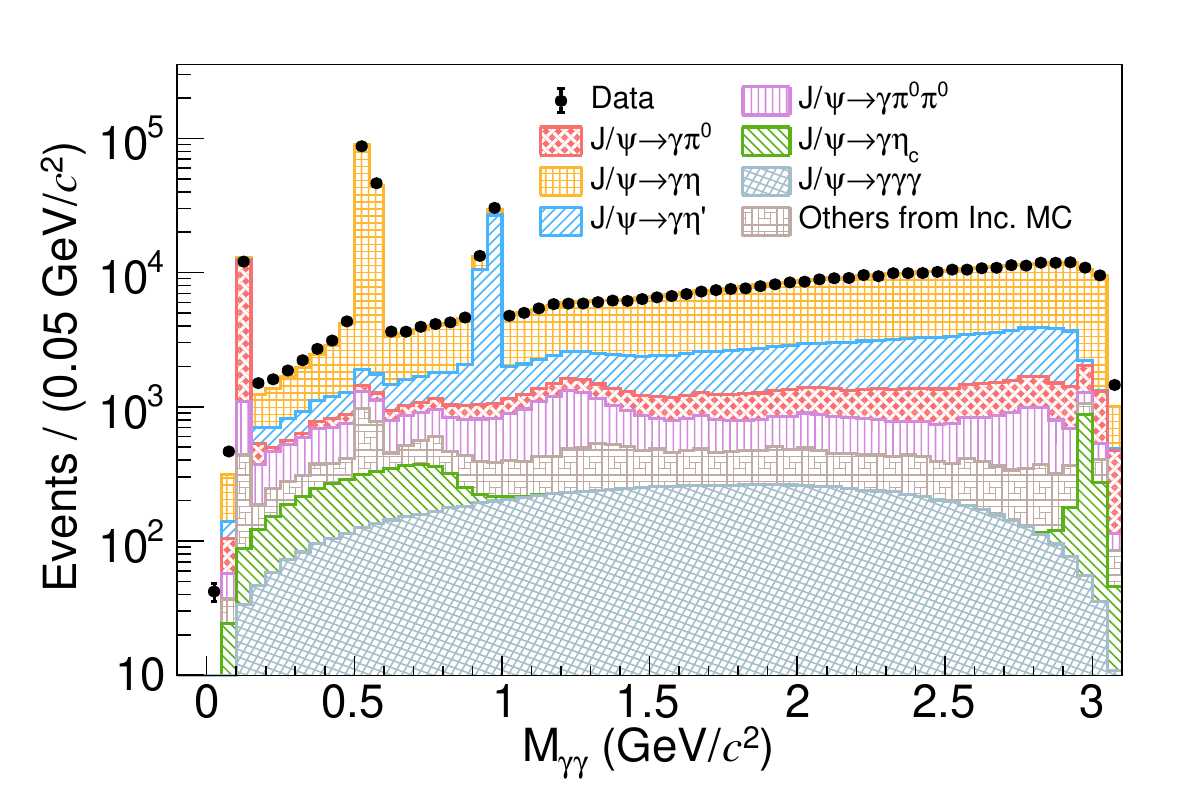}
    \end{overpic}
\end{center}
\vspace{-0.6cm}
\caption{The diphoton invariant-mass distributions for data and the MC-simulated backgrounds, which are normalized to the integrated luminosity of the data.}
\vspace{-0.2cm}
\label{fig:mgg}
\end{figure}

The two-photon invariant mass $\mgg$ distribution of the events selected by the above criteria is shown in Fig.~\ref{fig:mgg}. The MC simulation shows a rather good agreement with the data, corroborating the analysis procedure. There are three entries per event from all possible combinations of the three selected photons. The background is dominated by contributions from $J/\psi\rightarrow\gamma\pi^0$, $\gamma\eta$ and $\gamma\eta'$ decays. There is a small contribution from residual $J/\psi\rightarrow\gamma\pi^0\pi^0$ background with two soft photons peaking around the $\pi^0$ mass region, and a contamination from $ J/\psi\rightarrow\gamma f_2(1270)$ decays, followed by $f_2(1270)\rightarrow\pi^0\pi^0$. However, since the $f_2(1270)$ state has a large width and the contribution is relatively small, this contribution can be treated as non-peaking background and is well described by a polynomial function when extracting the signal yields.

A series of one-dimensional unbinned extended maximum-likelihood fits are performed to the $\mgg$ distribution to determine the signal yields with different ALP mass hypotheses in the mass range of $0.165\leq m_a\leq2.84\gevcc$.
The search step is $3\mevcc$ for $0.165\leq m_a<1.20\gevcc$ and $4\mevcc$ for other mass regions. A total of 674 mass hypotheses are probed.
The likelihood function is a combination of signal, non-peaking background, and peaking components of the $\pi^0$, $\eta$ and $\eta'$ mesons. 
Different $\mgg$ fit intervals are used for various $m_a$ hypotheses in order to handle the non-peaking and peaking backgrounds properly, which are listed in Table~\ref{tab:range}.

\begin{table}[H]
\centering
\vspace{-0.4cm}
\caption{The $M_{\gamma\gamma}$ fit intervals for various $m_a$ points.}
\vspace{-0.1cm}
\label{tab:range}
\begin{tabular}{cc}
\hline$m_a$ points ($\mbox{GeV}/c^2$) & $M_{\gamma\gamma}$ fit intervals ($\mbox{GeV}/c^2$)\\
\hline 0.165 - 0.35 & 0.06 - 0.45 \\
       0.35 - 0.75  & 0.25 - 0.85 \\
       0.75 - 1.20  & 0.65 - 1.30 \\
       1.20 - 2.84  & ($m_a-0.2$) - ($m_a+0.2$) \\
\hline
\end{tabular}
\end{table}
\vspace{-0.2cm}

\begin{figure}[H]
	\vspace{-1.3cm}
	\begin{center}
		\begin{overpic}[width=8.0cm]{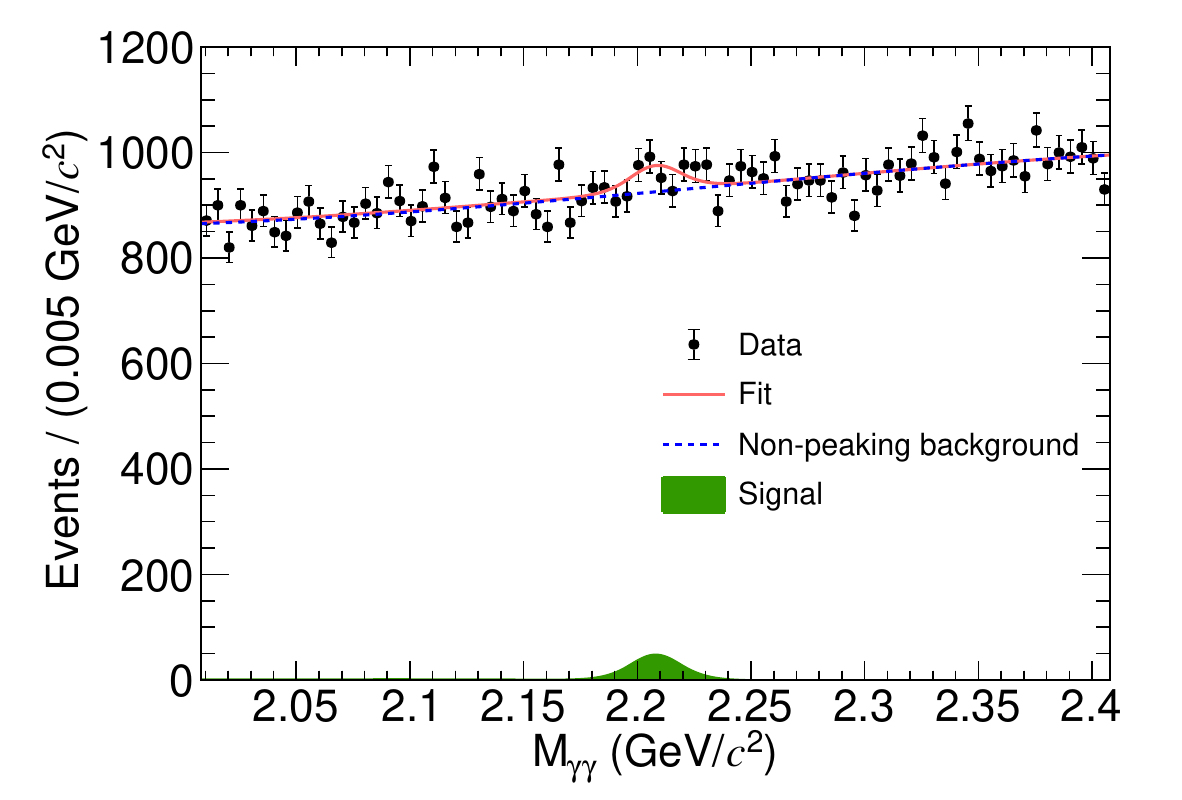}
			\put(20,58){\small{(b)}}
		\end{overpic}
		\vspace{-0.6cm}
		\caption{The $\mgg$ distribution for (a) $m_a=0.327\gevcc$ and (b) $m_a=2.208\gevcc$ with the fit results overlaid. The black dots with error bars are data, and the contribution of the non-peaking background is represented by a blue dashed curve. The green shaded region and the red solid curve represent the signal PDF and the total PDF, respectively. The inset of the figure (a) displays an enlargement of the $\mgg$ region between 0.25 and $0.4\gevcc$. The largest local significance is 2.6$\sigma$ at the $m_a=2.208\gevcc$ hypothesis.}
		\label{fig:fit}
	\end{center}
	\vspace{-0.6cm}
\end{figure}

The signal probability density function (PDF) is constructed from a peaking component from the ALP decay, and a combinatorial contribution from the other two combinations of photons. The peaking component is parameterized by the sum of two Crystal Ball (CB)~\cite{CB} functions with opposite-side tails with the same mean, resolution, and relative weight. These values and the detection efficiency are determined by fitting the signal MC samples. The resolution ranges from $6\mevcc$ near $m_a=0.165\gevcc$ to $11\mevcc$ near $m_a=2.2\gevcc$, and it decreases back to $7\mevcc$ near $m_a=2.84\gevcc$. The detection efficiency ranges from 30\% to 35\%. 
The combinatorial component of the signal is described by a smoothed kernel density estimation PDF~\cite{keys} obtained from the signal MC sample with the mass hypothesis $m_a$ closest to the fitted assumption. The ratio between the combinatorial part and the peaking component of the signal is determined for each fit interval. For any search point, the parameters of the CB functions, the ratio of the combinatorial part and the efficiency are interpolated between the mass points by a fit with a polynomial function.
The peaking background components at the $\pi^0$, $\eta$ and $\eta'$ masses are described by the sum of two CB functions using parameters determined from data. A fifth-order Chebyshev polynomial function is used to describe the non-peaking background for $0.75\leq m_a<1.20\gevcc$ and a third-order Chebyshev polynomial function is used for the remaining mass regions. 
The choice of the order of polynomial function is optimized through a spurious signal test by performing a signal-plus-background fit to the background-only $\mgg$ distribution~\cite{spurious}. The obtained signal yield (spurious signal) is required to be less than 30\% of its expected statistical uncertainty.
When performing the fit, the shapes of the signal and the peaking background PDFs are fixed, while the non-peaking background PDF shape and the yields of signal, peaking and non-peaking background events are free parameters. The $\mgg$ distributions for selected $m_a$ points and the fit results are shown in Fig.~\ref{fig:fit}.

The branching fractions of $J/\psi\rightarrow\gamma(\pi^0,\eta,\eta')\rightarrow\gamma\gamma\gamma$ are measured to validate the signal extraction procedure. The peaks are treated as signals, and the fitting procedure described above is applied to extract the number of peak events. After accounting for the contribution of peaking backgrounds, the results are found to be compatible with the published BESIII measurement~\cite{br_pseudo} within uncertainties.
\vspace{-0.2cm}

\section{Systematic uncertainties}

The systematic uncertainties are divided into two parts, which are additive and multiplicative in nature. The additive systematic uncertainty arises from the uncertainties in the signal parameterization and background modeling. The multiplicative systematic uncertainty includes contributions coming from MDC tracking, photon reconstruction, and selection criteria, which are related to the signal efficiency.

\begin{figure}[H]
	\vspace{-0.4cm}
	\begin{center}
		\begin{overpic}[width=8.0cm]{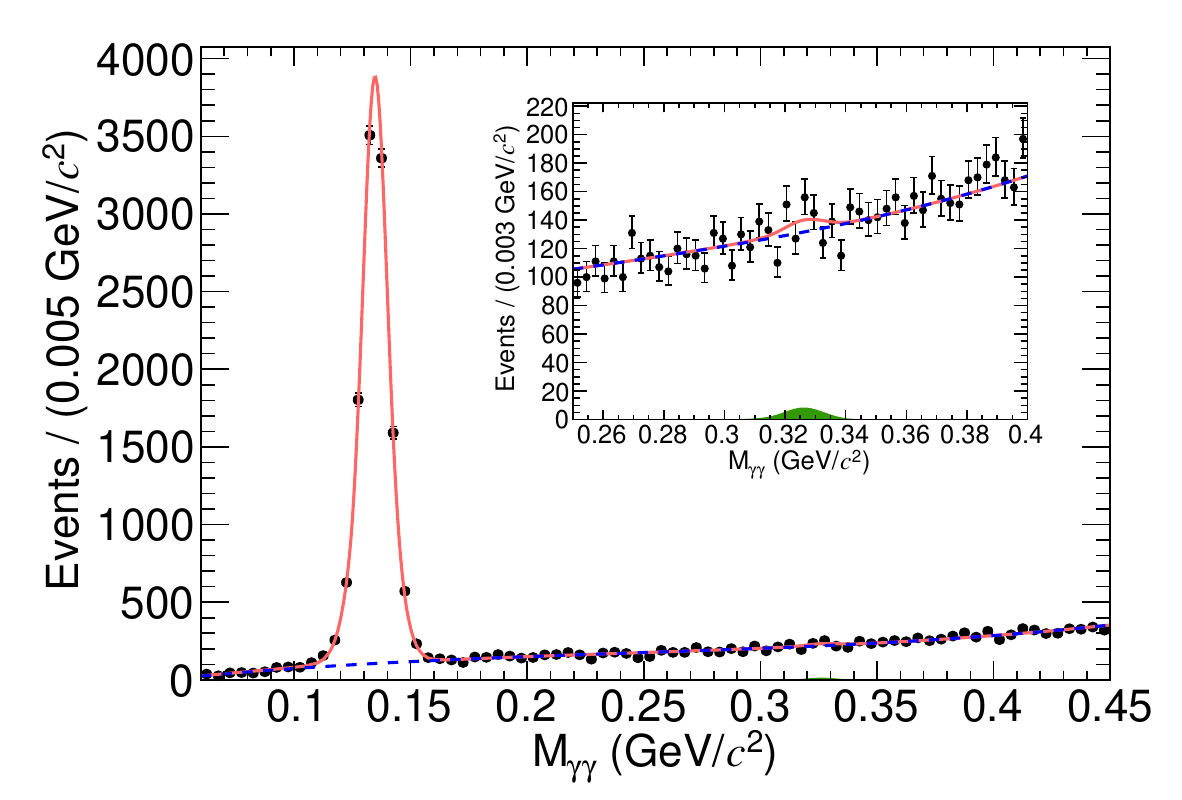}
			\put(20,58){\small{(a)}}
		\end{overpic}
	\end{center}
\end{figure}

The uncertainty due to signal and peaking background shapes is considered by performing the same fitting procedure with alternative signal and peaking background shapes. For the alternative signal shape, the parameters of the CB functions (except the mean value) and the ratio of the combinatorial part are obtained from the signal MC sample with the mass hypothesis $m_a$ closest to the search point. The alternative peaking background shape is modeled with the sum of a CB function and a Gaussian function. For each mass value $m_a$, the fit is performed three times in total with different methods, and the minimum significance value and the maximum upper limit are recorded.
The non-peaking background is described by polynomials. By performing the spurious signal test, the maximum of the absolute values of the spurious signal over the search ranges constitutes the related systematic uncertainty, which is the dominant source of systematic uncertainty in the analysis and is incorporated in the overall likelihood assuming a Gaussian distribution: 8.09 events for $0.165\leq m_a<0.35\gevcc$, 1.41 events for $0.35\leq m_a<0.75\gevcc$, 17.95 events for $0.75\leq m_a<1.20\gevcc$ and 39.93 events for $1.20\leq m_a\leq2.84\gevcc$.

The multiplicative systematic uncertainties are listed in Table~\ref{tab:multi}. The tracking efficiency of charged pions is  investigated using control samples of $J/\psi\rightarrow p\bar{p}\pi^+\pi^-$ decays~\cite{tracking}. The difference in tracking efficiencies between data and MC simulation is found to be 1.0\% per track, which is taken as the uncertainty for the tracking efficiency. The photon detection efficiency is studied with a clean sample of $J/\psi\rightarrow\rho^0\pi^0$ decays~\cite{photon}. The difference in detection efficiencies between data and the MC simulation is 1.0\% per photon. The systematic uncertainty due to the requirement on $M(\pi^+\pi^-)_{\rm recoil}$ is determined to be 0.3\% according to the study of $\psi(3686)\rightarrow\pi^+\pi^-J/\psi$, $J/\psi\rightarrow e^+e^-\mu^+\mu^-$ decays. The systematic uncertainties associated with the 4C kinematic fit, $E_{\rm other}$ and $N_{\rm trk}$ (number of charged tracks) requirements are studied with the control sample of $\psi(3686)\rightarrow\pi^+\pi^-J/\psi$, $J/\psi\rightarrow\gamma\eta$ decays and are determined to be 1.8\%, 1.4\% and 0.1\%, respectively. The uncertainty on the $\psi(3686)\rightarrow\pi^+\pi^-J/\psi$ branching fraction is taken from the PDG~\cite{pdg}, and the systematic uncertainty due to the number of $\psi(3686)$ events is determined to be 0.6\% by studying inclusive hadronic decays~\cite{number,number2}. The multiplicative systematic uncertainty is included in the overall likelihood as a Gaussian nuisance parameter with a width equal to the uncertainty.

\begin{table}[H]
\centering
\caption{The multiplicative systematic uncertainties. The total systematic uncertainty is obtained by adding all individual uncertainties in quadrature, assuming all sources to be independent.}
\vspace{-0.1cm}
\label{tab:multi}
\begin{tabular}{cc}
\hline Source                                         & Uncertainty (\%)\\
\hline MDC tracking                                   & 2.0 \\
Photon reconstruction                                 & 3.0 \\
$M(\pi^+\pi^-)_{\rm recoil}$ requirement              & 0.3 \\
4C kinematic fit                                      & 1.8 \\
$E_{\rm other}$ requirement                           & 1.4 \\
$N_{\rm trk}$ requirement                             & 0.1 \\
$\mathcal{B}(\psi(3686)\rightarrow\pi^+\pi^-J/\psi)$  & 0.9 \\
Number of $\psi(3686)$                                & 0.6 \\
\hline Total                                          & 4.4 \\
\hline
\end{tabular}
\end{table}
\vspace{-0.2cm}

\section{Result}

\begin{figure}[H]
\vspace{-0.4cm}
\begin{center}
    \begin{overpic}[width=8cm]{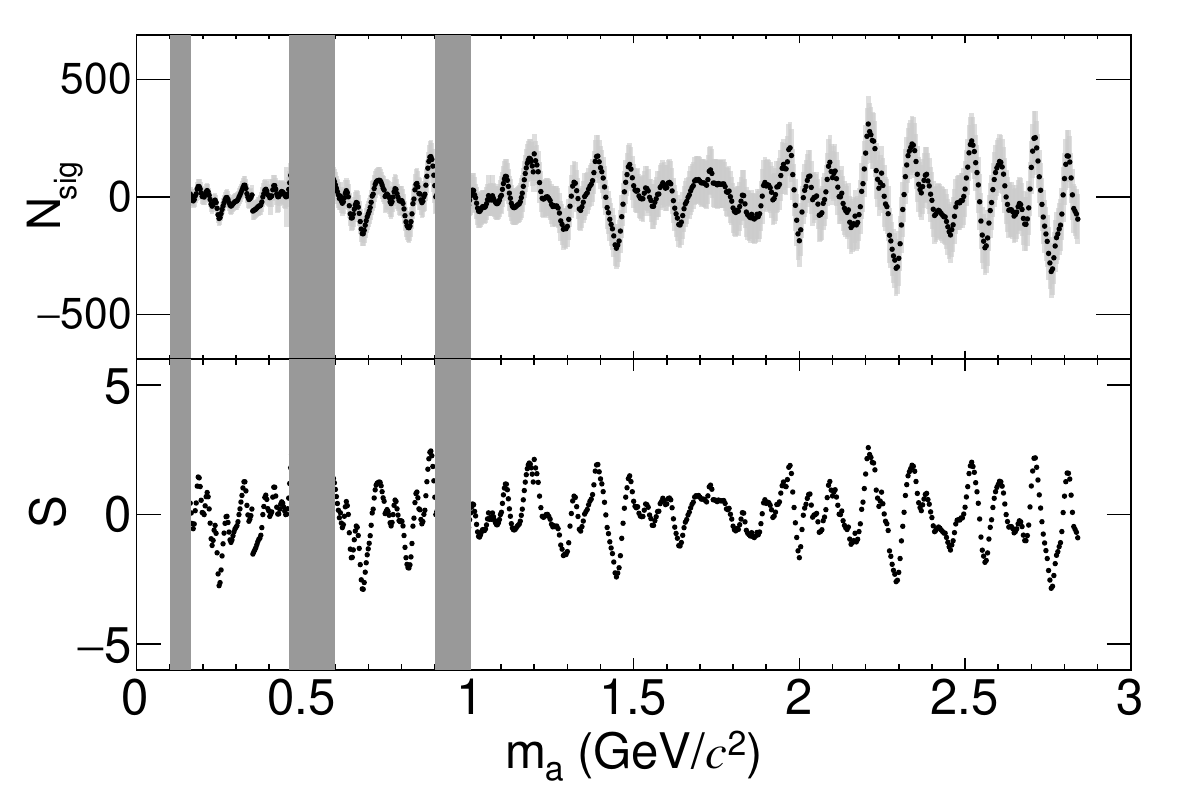}
    \end{overpic}
\end{center}
\vspace{-0.5cm}
\caption{The distribution of the signal yields $N_{\rm sig}$ (upper plot) and the local significance $\mathcal{S}$ (lower plot) obtained from the maximum-likelihood fits as a function of $m_a$. The vertical gray bands indicate the excluded regions close to the $\pi^0$, $\eta$, and $\eta'$ masses.}
\label{fig:nsig}
\vspace{-0.4cm}
\end{figure}

For each $m_a$ hypothesis, the local significance is determined by $\mathcal{S}=\mathrm{sign}(N_{\mathrm{sig}})\cdot\sqrt{2\ln(\mathcal{L}_{\mathrm{max}}/\mathcal{L}_0)}$, where $\mathcal{L}_{\mathrm{max}}$ and $\mathcal{L}_0$ are the likelihood values with and without the signal hypothesis included in the fit, respectively. The fitted signal yields $N_{\rm sig}$ and the corresponding significances are shown in Fig.~\ref{fig:nsig}. The largest local significance of $2.6\sigma$ is observed near $m_a=2.208\gevcc$, consistent with the null hypothesis. The fit results for $m_a=2.208\gevcc$ are shown in Fig.~\ref{fig:fit}(b).

Since no significant ALP signal is observed, we compute 95\% confidence level (CL) upper limits on $\mathcal{B}(J/\psi\rightarrow\gamma a)$ as a function of $m_a$ using a one-sided frequentist profile-likelihood method~\cite{CLs}. Then the branching fraction limit is converted to the ALP-photon coupling limit using~\cite{Merlo}
\vspace{-0.3cm}

\begin{equation}
g_{a\gamma\gamma}=\sqrt{\frac{\mathcal{B}(J/\psi\rightarrow\gamma a)}{\mathcal{B}(J/\psi\rightarrow e^+e^-)}(1-\frac{m_a^2}{m_{J/\psi}^2})^{-3}\frac{32\pi\alpha_{\mathrm{em}}}{m_{J/\psi}^2}},
\end{equation} 
where $\mathcal{B}(J/\psi\rightarrow e^+e^-)=(5.971\pm0.032)\%$ is the world average value from the PDG~\cite{pdg} and $\alpha_{\mathrm{em}}$ is the electromagnetic coupling. An additional 0.5\% uncertainty arising from the knowledge of  $\mathcal{B}(J/\psi\rightarrow e^+e^-)$ is included when converting $\mathcal{B}(J/\psi\rightarrow\gamma a)$ to $g_{a\gamma\gamma}$.

\begin{figure}[H]
\vspace{-0.2cm}
\begin{center}
    \begin{overpic}[width=8cm]{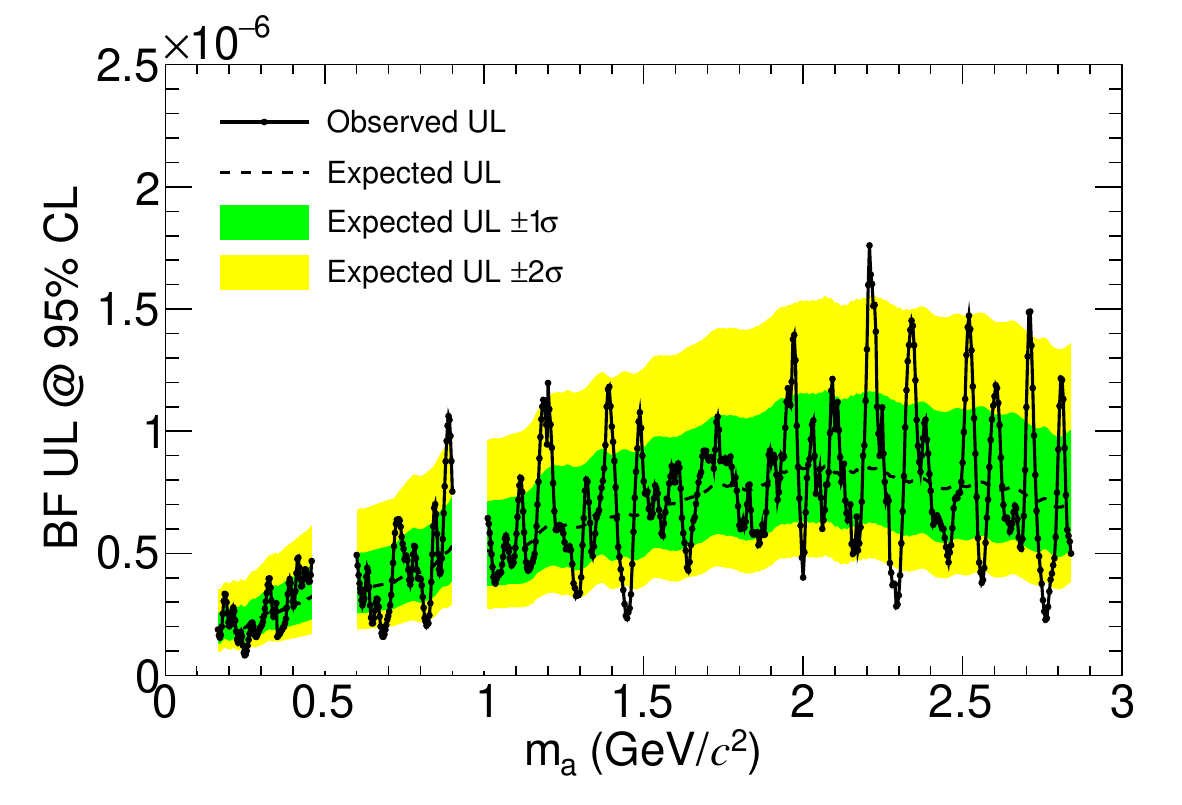}
    \end{overpic}
\end{center}
\vspace{-0.6cm}
\caption{Expected and observed upper limits at 95\% CL on $\mathcal{B}(J/\psi\rightarrow\gamma a)$. The black curve is for the data, the black dashed curve represents the expected values and the green (yellow) band represents the $\pm1\sigma$ ($\pm2\sigma$) region.}
\label{fig:UL}
\vspace{-0.2cm}
\end{figure}

\begin{figure}[H]
\vspace{-0.4cm}
\begin{center}
    \begin{overpic}[width=8cm]{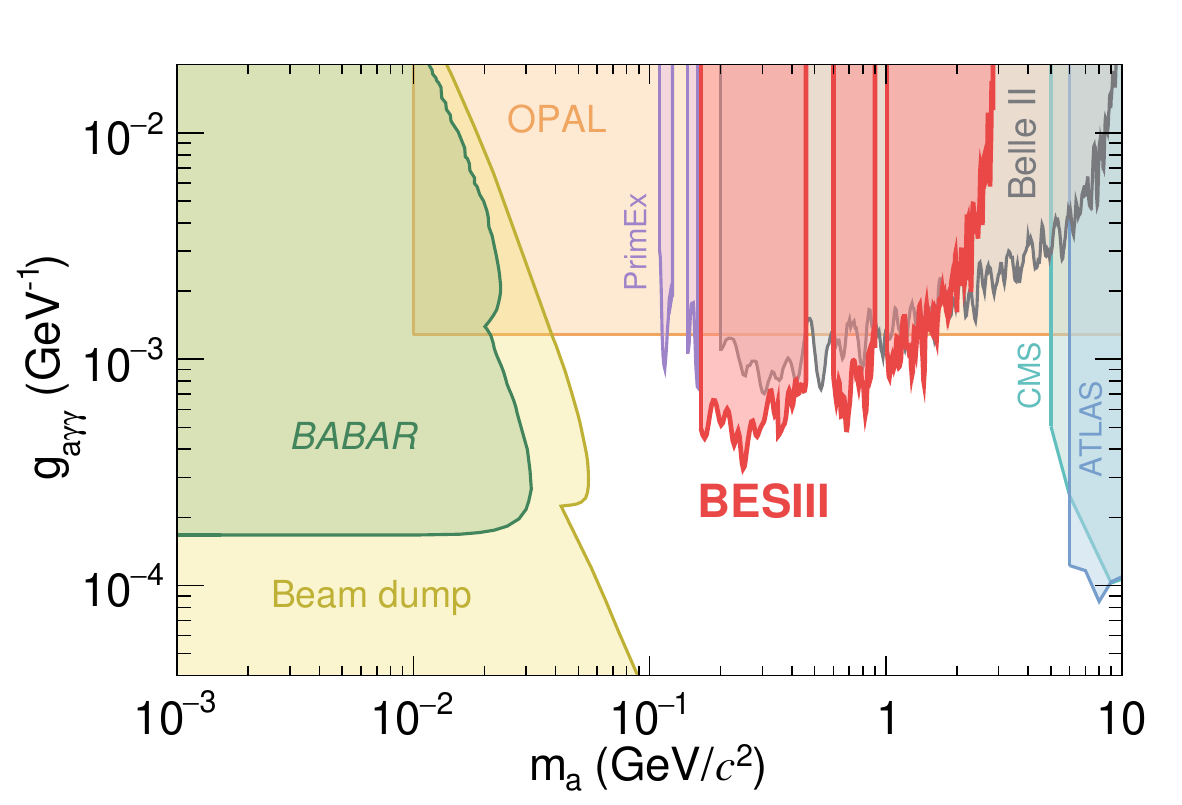}
    \end{overpic}
\end{center}
\vspace{-0.5cm}
\caption{Exclusion limits at 95\% CL in the ALP-photon coupling $g_{a\gamma\gamma}$ versus ALP mass $m_a$ plane obtained from this analysis. The constraints from \emph{BABAR}~\cite{BaBar}, OPAL~\cite{OPAL}, Belle~II~\cite{Belle}, CMS~\cite{CMS}, ATLAS~\cite{ATLAS}, PrimEx~\cite{PrimEx}, and beam-dump experiments~\cite{Beam_dump,NA64} are also shown. All measurements assume a 100\% ALP decay branching fraction into photons.}
\label{fig:coupling}
\end{figure}

The expected and observed upper limits at 95\% CL on $\mathcal{B}(J/\psi\rightarrow\gamma a)$ are shown in Fig.~\ref{fig:UL}. The observed limits range from $8.3\times10^{-8}$ to $1.8\times10^{-6}$ in the ALP mass region of $0.165\leq m_a\leq2.84\gevcc$. The exclusion limits in the ALP-photon coupling $g_{a\gamma\gamma}$ versus ALP mass $m_a$ plane obtained from this analysis are shown in Fig.~\ref{fig:coupling}, together with the constraints of other experiments. Our limits exclude the region in the ALP-photon coupling range $g_{a\gamma\gamma}>3\times10^{-4}\,\mbox{GeV}^{-1}$ for an ALP mass $m_a$ around $0.25\gevcc$, with an improvement by a factor of 2-3 over the previous Belle~II measurement~\cite{Belle}. In addition, the constraints on the ALP-photon coupling are the most stringent to date for $0.165\leq m_a\leq1.468\gevcc$.

\vspace{-0.4cm}

\section{Summary}

Based on a data sample of $(2.71\pm0.01)\times10^9$ $\psi(3686)$ events collected by the BESIII detector, we search for ALPs decaying into two photons produced in $J/\psi$ radiative decays using the $\psi(3686)\rightarrow\pi^+\pi^-J/\psi$ process. No significant ALP signal is observed and we set 95\% CL upper limits on the branching fraction of the decay $J/\psi\rightarrow\gamma a$ and the ALP-photon coupling $g_{a\gamma\gamma}$. The observed limits on $\mathcal{B}(J/\psi\rightarrow\gamma a)$ range from $8.3\times10^{-8}$ to $1.8\times10^{-6}$ in the ALP mass region of $0.165\leq m_a\leq2.84\gevcc$, and the exclusion limits on the ALP-photon coupling are the most stringent to date for $0.165\leq m_a\leq1.468\gevcc$.

\vspace{-0.2cm}

\section*{Acknowledgement}

The BESIII collaboration thanks the staff of BEPCII and the IHEP computing center for their strong support. We thank Jia Liu for the useful suggestions. This work is supported in part by National Key R\&D Program of China under Contracts Nos. 2020YFA0406400, 2020YFA0406300; National Natural Science Foundation of China (NSFC) under Contracts Nos. 11635010, 11735014, 11835012, 11935015, 11935016, 11935018, 11961141012, 12022510, 12025502, 12035009, 12035013, 12061131003, 12192260, 12192261, 12192262, 12192263, 12192264, 12192265; the Chinese Academy of Sciences (CAS) Large-Scale Scientific Facility Program; the CAS Center for Excellence in Particle Physics (CCEPP); Joint Large-Scale Scientific Facility Funds of the NSFC and CAS under Contract No. U1832207; CAS Key Research Program of Frontier Sciences under Contracts Nos. QYZDJ-SSW-SLH003, QYZDJ-SSW-SLH040; 100 Talents Program of CAS; The Institute of Nuclear and Particle Physics (INPAC) and Shanghai Key Laboratory for Particle Physics and Cosmology; ERC under Contract No. 758462; European Union's Horizon 2020 research and innovation programme under Marie Sklodowska-Curie grant agreement under Contract No. 894790; German Research Foundation DFG under Contracts Nos. 443159800, 455635585, Collaborative Research Center CRC 1044, FOR5327, GRK 2149; Istituto Nazionale di Fisica Nucleare, Italy; Ministry of Development of Turkey under Contract No. DPT2006K-120470; National Research Foundation of Korea under Contract No. NRF-2022R1A2C1092335; National Science and Technology fund; National Science Research and Innovation Fund (NSRF) via the Program Management Unit for Human Resources \& Institutional Development, Research and Innovation under Contract No. B16F640076; Olle Engkvist Foundation under Contract No. 200-0605; Polish National Science Centre under Contract No. 2019/35/O/ST2/02907; STFC (United Kingdom); Suranaree University of Technology (SUT), Thailand Science Research and Innovation (TSRI), and National Science Research and Innovation Fund (NSRF) under Contract No. 160355; The Royal Society, UK under Contracts Nos. DH140054, DH160214; The Swedish Research Council; U.S. Department of Energy under Contract No. DE-FG02-05ER41374

\end{multicols}


\begin{thebibliography}{00}

\bibitem{axion_1}
\href{https://journals.aps.org/prl/abstract/10.1103/PhysRevLett.38.1440}{
R. D. Peccei and H. R. Quinn,
Phys. Rev. Lett. \textbf{38}, 1440 (1977).} 

\bibitem{axion_2}
\href{https://journals.aps.org/prd/abstract/10.1103/PhysRevD.16.1791}{
R. D. Peccei and H. R. Quinn,
Phys. Rev. D \textbf{16}, 1791 (1977).}

\bibitem{axion_3}
\href{https://journals.aps.org/prl/abstract/10.1103/PhysRevLett.40.223}{
S. Weinberg,
Phys. Rev. Lett. \textbf{40}, 223 (1978).} 

\bibitem{axion_4}
\href{https://journals.aps.org/prl/abstract/10.1103/PhysRevLett.40.279}{
F. Wilczek,
Phys. Rev. Lett. \textbf{40}, 279 (1978).} 

\bibitem{hierarchy}
\href{https://journals.aps.org/prl/abstract/10.1103/PhysRevLett.115.221801}{
P. W. Graham, D. E. Kaplan and S. Rajendran, 
Phys. Rev. Lett. \textbf{115}, 221801 (2015).}

\bibitem{string}
\href{https://iopscience.iop.org/article/10.1088/1742-6596/485/1/012013}{
A. Ringwald,
J. Phys. Conf. Ser. \textbf{485}, 012013 (2014).}

\bibitem{Higgs}
\href{https://doi.org/10.1016/j.physrep.2012.02.002}{
G. C. Branco {\it et al.},
Phys. Rept. \textbf{516}, 1 (2012).}

\bibitem{dark_1}
\href{https://journals.aps.org/prd/abstract/10.1103/PhysRevD.83.115009}{
M. Freytsis and Z. Ligeti,
Phys. Rev. D \textbf{83}, 115009 (2011).}

\bibitem{dark_2}
\href{https://doi.org/10.1016/0370-2693(83)90637-8}{
J. Preskill, M. B. Wise and F. Wilczek,
Phys. Lett. B \textbf{120}, 127 (1983).}

\bibitem{dark_3}
\href{https://doi.org/10.1016/0370-2693(83)90638-X}{
L. F. Abbott and P. Sikivie,
Phys. Lett. B \textbf{120}, 133 (1983).}

\bibitem{dark_4}
\href{https://doi.org/10.1016/0370-2693(83)90639-1}{
M. Dine and W. Fischler,
Phys. Lett. B \textbf{120}, 137 (1983).}

\bibitem{cosmo_1}
\href{https://www.annualreviews.org/doi/abs/10.1146/annurev-nucl-102014-022120}{
P. W. Graham {\it et al.},
Rev. Nucl. Part. Sci. \textbf{65}, 485 (2015).}

\bibitem{cosmo_2}
\href{https://iopscience.iop.org/article/10.1088/1475-7516/2012/02/032}{
D. Cadamuro and J. Redondo,
JCAP \textbf{02}, 032 (2012).}

\bibitem{PrimEx}
\href{https://journals.aps.org/prl/abstract/10.1103/PhysRevLett.123.071801}{
D. Aloni {\it et al.},
Phys. Rev. Lett. \textbf{123}, 071801 (2019).}

\bibitem{Beam_dump}
\href{https://link.springer.com/article/10.1007/JHEP02(2016)018}{
B. Döbrich {\it et al.},
JHEP \textbf{02}, 018 (2016).}

\bibitem{NA64}
\href{https://journals.aps.org/prl/abstract/10.1103/PhysRevLett.125.081801}{
D. Banerjee {\it et al.} [NA64 Collaboration],
Phys. Rev. Lett. \textbf{125}, 081801 (2020).}

\bibitem{meson1}
\href{https://journals.aps.org/prd/abstract/10.1103/PhysRevD.101.112005}{
M. Ablikim {\it et al.} [BESIII Collaboration],
Phys. Rev. D \textbf{101}, 112005 (2020).}

\bibitem{meson2}
\href{https://journals.aps.org/prl/abstract/10.1103/PhysRevLett.122.011801}{
I. S. Seong {\it et al.} [Belle Collaboration],
Phys. Rev. Lett. \textbf{122}, 011801 (2019).}

\bibitem{BaBar}
\href{https://link.springer.com/article/10.1007/JHEP12(2017)094}{
M. J. Dolan {\it et al.},
JHEP \textbf{12}, 094 (2017);} 
\href{https://link.springer.com/article/10.1007/JHEP03(2021)190}{
Erratum: [JHEP \textbf{03}, 190 (2021)].}

\bibitem{OPAL}
\href{https://journals.aps.org/prl/abstract/10.1103/PhysRevLett.118.171801}{
S. Knapen {\it et al.},
Phys. Rev. Lett. \textbf{118}, 171801 (2017).}

\bibitem{L3}
\href{https://doi.org/10.1016/0370-2693(95)01612-T}{
M. Acciarri {\it et al.} [L3 Collaboration],
Phys. Lett. B \textbf{345}, 609 (1995).}

\bibitem{Belle}
\href{https://journals.aps.org/prl/abstract/10.1103/PhysRevLett.125.161806}{
F. Abudinén {\it et al.} [Belle II Collaboration], 
Phys. Rev. Lett. \textbf{125}, 161806 (2020).}

\bibitem{CMS}
\href{https://doi.org/10.1016/j.physletb.2019.134826}{
A. M. Sirunyan {\it et al.} [CMS Collaboration],
Phys. Lett. B \textbf{797}, 134826 (2019).}

\bibitem{ATLAS}
\href{https://link.springer.com/article/10.1007/JHEP03(2021)243}{
G. Aad {\it et al.} [ATLAS Collaboration],
JHEP \textbf{03}, 243 (2021);} 
\href{https://link.springer.com/article/10.1007/JHEP11(2021)050}{
Erratum: [JHEP \textbf{11}, 050 (2021)].}

\bibitem{collider}
\href{https://arxiv.org/abs/2102.08971}{
D. d’Enterria, 
Workshop on Feebly Interacting Particles (2021), arXiv:2102.08971.}

\bibitem{pp1}
\href{https://doi.org/10.1103/PhysRevLett.113.171801}{
G. Aad {\it et al.} [ATLAS Collaboration],
Phys. Rev. Lett. \textbf{113}, 171801 (2014).}

\bibitem{pp2}
\href{https://link.springer.com/article/10.1140/epjc/s10052-016-4034-8}{
G. Aad {\it et al.} [ATLAS Collaboration],
Eur. Phys. J. C \textbf{76}, 210 (2016).}

\bibitem{LHC}
\href{https://doi.org/10.1016/j.physletb.2015.12.037}{
J. Jaeckel, M. Spannowsky,
Phys. Lett. B \textbf{753}, 482 (2016).}

\bibitem{Merlo}
\href{https://link.springer.com/article/10.1007/JHEP06(2019)091}{
L. Merlo {\it et al.},
JHEP \textbf{06}, 091 (2019).}

\bibitem{Ablikim:2009aa}
\href{https://doi.org/10.1016/j.nima.2009.12.050}{
M.~Ablikim {\it et al.} [BESIII Collaboration],
Nucl.\ Instrum.\ Meth.\ A {\bf 614}, 345 (2010).}

\bibitem{Yu:IPAC2016-TUYA01}
\href{https://accelconf.web.cern.ch/ipac2016/papers/tuya01.pdf}{
C.~H.~Yu {\it et al.},
Proceedings of IPAC2016, Busan, Korea, 2016, doi:10.18429/JACoW-IPAC2016-TUYA01.}

\bibitem{Ablikim:2019hff}
\href{https://iopscience.iop.org/article/10.1088/1674-1137/44/4/040001}{
M.~Ablikim {\it et al.} [BESIII Collaboration],
Chin. Phys. C {\bf 44}, 040001 (2020).}

\bibitem{etof}
\href{https://link.springer.com/article/10.1007/s41605-017-0014-2}{
X.~Li {\it et al.}, 
Radiat. Detect. Technol. Methods {\bf 1}, 13 (2017);}
\href{https://link.springer.com/article/10.1007/s41605-017-0012-4}{
Y.~X.~Guo {\it et al.}, 
Radiat. Detect. Technol. Methods {\bf 1}, 15 (2017);}
\href{https://doi.org/10.1016/j.nima.2019.163053}{
P.~Cao {\it et al.}, 
Nucl.\ Instrum.\ Meth.\ A {\bf 953}, 163053 (2020).}

\bibitem{number}
\href{https://iopscience.iop.org/article/10.1088/1674-1137/37/6/063001}{
M. Ablikim {\it et al.} [BESIII Collaboration], 
Chin. Phys. C \textbf{37}, 063001 (2013).}

\bibitem{number2}
\href{https://iopscience.iop.org/article/10.1088/1674-1137/42/2/023001}{
M. Ablikim {\it et al.} [BESIII Collaboration], 
Chin. Phys. C \textbf{42}, 023001 (2018).}

\bibitem{geant4}
\href{https://doi.org/10.1016/S0168-9002(03)01368-8}{
S.~Agostinelli {\it et al.} [GEANT4 Collaboration],
Nucl.\ Instrum.\ Meth.\ A {\bf 506}, 250 (2003).}

\bibitem{ref:kkmc}
\href{https://journals.aps.org/prd/abstract/10.1103/PhysRevD.63.113009}{
S.~Jadach, B.~F.~L.~Ward and Z.~Was,
Phys.\ Rev.\ D {\bf 63}, 113009 (2001);}
\href{https://doi.org/10.1016/S0010-4655(00)00048-5}{
Comput.\ Phys.\ Commun.\  {\bf 130}, 260 (2000).}

\bibitem{ref:evtgen}
\href{https://doi.org/10.1016/S0168-9002(01)00089-4}{
D.~J.~Lange,
Nucl.\ Instrum.\ Meth.\ A {\bf 462}, 152 (2001);}
\href{https://iopscience.iop.org/article/10.1088/1674-1137/32/8/001}{  
R.~G.~Ping,
Chin. Phys. C {\bf 32}, 599 (2008).}

\bibitem{pdg}
\href{https://doi.org/10.1093/ptep/ptac097}{
R. L. Workman {\it et al.} [Particle Data Group], 
Prog. Theor. Exp. Phys. {\bf 2022}, 083C01 (2022).}

\bibitem{ref:lundcharm}
\href{https://journals.aps.org/prd/abstract/10.1103/PhysRevD.62.034003}{
J.~C.~Chen {\it et al.},
Phys.\ Rev.\ D {\bf 62}, 034003 (2000);}
\href{https://iopscience.iop.org/article/10.1088/0256-307X/31/6/061301}{  R.~L.~Yang, R.~G.~Ping and H.~Chen,
Chin.\ Phys.\ Lett.\  {\bf 31}, 061301 (2014).}

\bibitem{photos}
\href{https://doi.org/10.1016/0370-2693(93)90062-M}{ 
E.~Richter-Was,
Phys.\ Lett.\ B {\bf 303}, 163 (1993).}

\bibitem{jpipi}
\href{https://journals.aps.org/prd/abstract/10.1103/PhysRevD.62.032002}{ 
J. Z. Bai, {\it et al.} [BES Collaboration],
Phys.\ Rev.\ D {\bf 62}, 032002 (2000).}

\bibitem{topoana}
\href{https://doi.org/10.1016/j.cpc.2020.107540}{
X. Y. Zhou {\it et al.}, Comput. Phys. Commun. {\bf 258}, 107540 (2021).}

\bibitem{pwa_gpipi}
\href{https://journals.aps.org/prd/abstract/10.1103/PhysRevD.92.052003}{ 
M.~Ablikim {\it et al.} [BESIII Collaboration],
Phys.\ Rev.\ D {\bf 92}, 052003 (2015);}
\href{https://journals.aps.org/prd/abstract/10.1103/PhysRevD.93.039906}{
Erratum: [Phys.\ Rev.\ D {\bf 93}, 039906 (2016)].}

\bibitem{punzi}
\href{https://arxiv.org/abs/physics/0308063}{ 
G. Punzi,
eConf {\bf C030908}, MODT002 (2003), arXiv:physics/038063.}

\bibitem{CB}
\href{https://doi.org/10.1016/j.nima.2016.05.010}{ 
J. H. Cheng {\it et al.},
Nucl. Instrum. Methods Phys. Res., Sect. A {\bf 827}, 165 (2016).}

\bibitem{keys}
\href{https://doi.org/10.1016/S0010-4655(00)00243-5}{ 
K.~S.~Cranmer,
Comput.\ Phys.\ Commun.\  {\bf 136}, 198 (2001).}

\bibitem{spurious}
\href{https://cds.cern.ch/record/2743717}{
ATLAS Collaboration, 
{\it Recommendations for the Modeling of Smooth Backgrounds}, ATL-PHYS-PUB-2020-028, 2020.}

\bibitem{br_pseudo}
\href{https://journals.aps.org/prd/abstract/10.1103/PhysRevD.97.072014}{ 
M.~Ablikim {\it et al.} [BESIII Collaboration],
Phys. Rev. D {\bf 97}, 072014 (2018).}

\bibitem{tracking}
\href{https://journals.aps.org/prd/abstract/10.1103/PhysRevD.83.112005}{ 
M. Ablikim {\it et al.} [BESIII Collaboration], 
Phys. Rev. D \textbf{83}, 112005 (2011).}

\bibitem{photon}
\href{https://journals.aps.org/prd/abstract/10.1103/PhysRevD.81.052005}{ 
M. Ablikim {\it et al.} [BESIII Collaboration], 
Phys. Rev. D \textbf{81}, 052005 (2010).}

\bibitem{CLs}
\href{https://link.springer.com/article/10.1140/epjc/s10052-011-1554-0}{ 
G. Cowan {\it et al.},
Eur. Phys. J. C \textbf{71}, 1554 (2011);} 
\href{https://link.springer.com/article/10.1140/epjc/s10052-013-2501-z}{[Erratum: Eur. Phys. J. C \textbf{73}, 2501 (2013)].}

\end{thebibliography}
\end{document}